\documentstyle[twocolumn,prl,aps]{revtex}

\input epsf.sty

\begin{document}
\draft
\title{ 
Bose-Einstein condensation with                                                                  internal degrees of freedom in alkali atom gases
}

\author{ Tetsuo Ohmi$^{1\ast}$ and  Kazushige Machida$^{2\ast\ast}$}
\address{
$^{1}$Department of Physics, Kyoto University, Kyoto 
606-8502, Japan\\
$^{2}$Department of Physics, Okayama University,
         Okayama 700-8530, Japan}

\maketitle

\begin{abstract}
The Bogoliubov theory is extended to a Bose-Einstein condensation with internal degrees of freedom, realized recently in $^{23}$Na gases where several hyperfine states
are simultaneously cooled optically. Starting with a Hamiltonian constructed from general gauge and spin rotation symmetry principles fundamental equations for condensate are derived. The ground state where time reversal symmetry is broken in some case and low-lying collective modes, e.g. spin and density wave modes, are discussed. Novel vortex as a topological defect can be created experimentally. 
\end{abstract}


Much attention has been focused on Bose-Einstein 
condensation (BEC) since its realization in alkali atomic gases such as $^{87}$Rb\cite{cornell}, $^{23}$Na\cite{ketterle} and $^{7}$Li\cite{hulet}.
Furious experimental\cite{review1} and theoretical\cite{review2} investigations have
been done to reveal various novel phenomena of BEC which have not been attainable in another BEC system, that, is, superfluid $^{4}$He. One of the most notable differences
in these two systems lies in their mutual interaction either very weak in the former or very strong in the latter. This enables us to construct a microscopic theory for BEC in the present system from first principles. In fact a mean-field theory  or Hartree-Fock-Bogoliubov framework due to Bogoliubov\cite{bogoliubov}, Gross\cite{gross}, Pitaevskii\cite{pitaevskii} and others has been quite successful in explaining fundamental physics of BEC realized in magnetically trapped atomic gases such as condensate fraction, transition temperature, or low lying collective modes. We only feed the s-wave scattering length $a$ into these microscopic theories as a material parameter, which itself is known rather accurately\cite{review1,review2}.

Most experiments for BEC so far were performed by the magnetic traps which necessarily select one of several possible atomic ground states, or the so-called weak field seeking state, such as $F_z$=-1 among $F_z=\pm1,0$ of $F=1$ in the $^{23}$Na case, where the spin degrees of freedom are ``frozen". Recently Stamper-Kurn {\it et al.}\cite{kurn} have succeeded in cooling $^{23}$Na atoms purely optically in an optical dipole trap and achieved BEC, where the three substates $F_z=\pm1,0$ are simultaneously ``Bose-condensed". They demonstrate it by a ``Stern-Gerlach" experiment: The $F=1$ magnetic three sublevels are all visible as a split image by passing it through a field gradient. That is just remarkable.
This opens an interesting possibility to explore BEC with internal degrees of freedom where not only the gauge symmetry $U(1)$ but also ``spin" symmetry $SO(3)$ for the $F=1$ case are involved, a situation similar to the superfluid $^3$He problem where the orbital ($l=1$) and spin ($s=1$) degrees of freedom give rise to a $3\times3\times2$ dimensional manifold of complex order parameter space\cite{volhart}.

The purposes of this paper are (1) to establish fundamental equations for describing such a situation, leading to an extended Gross-Pitaevskii equation, (2) to examine low-lying collective modes in order to extract basic physical properties of the ground state and finally (3) to point out several interesting possible topological excitations or textural spatial structures under special circumstances, e.g. when releasing BEC from an optically plugged quadrupole magnetic trap, a novel vortex
may be created experimentally. So far no one has succeeded in observing a vortex even for one-component BEC although there have been several theoretical discussions\cite{vortex}. We will try to demonstrate that the present vectorial BEC system provides us a new horizon and fertile field to further explore\cite{siggia}.

In order to make our argument clear-cut, we consider a case described by a constituent atom with the hyperfine state $|F|=1$ ($F_z=\pm1,0$) (an extension of the theory to other cases is easy and trivial) where the order parameter or the Bose condensate is characterized by three components:
$\Psi_1, \Psi _0,\Psi_{-1}$ corresponding to each sublevel. Instead of choosing the basis set $|1>$, $|0>$, $|-1>$ where the quantization axis is chosen along the $z$-axis, 
we introduce the basis set $|x>, |y>, |z>$ defined by $S_i|i>=0 \ \ (i=x,y,z)$ where $S_i$ is the $i$-th component of the spin operator ($|S|=1$). The order parameter is now expressed by a three-dimensional vector ${\vec \Psi}=(\Psi_x,\Psi_y,\Psi_z)$ on this basis set which behaves as a vector under spin space rotation.

The Hamiltonian density can be constructed which should be invariant under spin space rotation and gauge transformation ($\hbar=1$):

\begin{eqnarray}
H=\psi_{\alpha}^{\dagger}(-\frac{\nabla^2} {2m}-\mu)\psi_{\alpha}+\frac{1}{2}g_1|\psi_{\alpha}|^4
\nonumber\\
+{1\over  2}g_2|\psi^2_{\alpha}|^2+i\varepsilon_{\alpha\beta\gamma}\omega_{L\gamma}\psi_{\alpha}^{\dagger}\psi_{\beta}
\end{eqnarray}
in terms of the creation and annihilation operators $\psi_{\alpha}^{\dagger},\psi_{\alpha}$ with the $\alpha$-th species Bose particle ($\alpha=x,y,z$).
The Larmor frequency in a vectorial notation $ {\vec \omega}_L=\gamma_{\mu} {\bf B}$ ($\gamma_{\mu}$: the gyromagnetic ratio and ${\bf B}$: the external magnetic field). The others are the standard notation (Repeated index is summed). The two kinds of the interaction terms $g_1$ and $g_2$ are characteristic  of the multi-component order parameters which represent the spin degrees of freedom of the condensate. These are physically originated from the interactions:

\begin{equation}
{1\over 2}g_n{\hat n}\cdot{\hat n}+{1\over 2}g_s{\bf {\hat S}}\cdot{\bf {\hat S}}
\end{equation}
where ${\hat n}=\psi_{\alpha}^{\dagger}\psi_{\alpha}$ and 
${\bf {\hat S}}_i=\psi_{\alpha}^{\dagger}S_i\psi_{\alpha}$ with $S_i$  being a $3\times3$ spin matrix.
It is easy to confirm the relations: $g_1=g_n+g_s$ and $g_2=-g_s$. Thus the $g_2$ term in (1) comes from the $F$-spin flipping process. It should be noted here that the precise values of $g_1$ and $g_2$ are neither known for each atomic species of interest (Na, Rb and Li) and nor determined for each hyperfine state ($|F|=1,2,\cdots$) to our knowledge. However, according to a recent spectroscopic experiment by Abraham {\it et al.}\cite{abraham} on $^7$Li (Boson) which determines the singlet ($a_s$) and triplet ($a_t$) scattering lengths ($a_s=33\pm2$ and $a_t=-27.6\pm0.5$ in units of the Bohr length), it is conceivable in general that the spin flipping interaction $g_2$ is comparable to $g_1$ and could be either positive or negative. For the present BEC to be stable it is necessary that $g_1+g_2>0$ as seen shortly.

Following the standard procedure~\cite{fetter} which leads to the Gross-Pitaevskii equation and starting with the Hamiltonian (1), we can derive the coupled equations for the condensate wave functions $\Psi_{\alpha}$

\begin{eqnarray}
i{\partial\over\partial t}\Psi_{\alpha}=(-\frac{\nabla^2} {2m}-\mu)\Psi_{\alpha}+g_1\Psi_{\beta}^{\ast}\Psi_{\beta}\Psi_{\alpha}
\nonumber\\
+g_2\Psi_{\beta}\Psi_{\beta}\Psi_{\alpha}^{\ast}+i\varepsilon_{\alpha\beta\gamma}\omega_{L\gamma}\Psi_{\beta}
\end{eqnarray}
where in the equation of motion for $\psi_{\alpha}$ the ground state average $<\psi_{\alpha}>$ is replaced by $\Psi_{\alpha}$ and the approximation such as $<\psi^{\dagger}\psi\psi>\sim\Psi^{\ast}\Psi\Psi$ is introduced. This is a basic set of equations of a non-linear coupled Shr$\ddot{\rm o}$dinger type.

When the system is uniform and infinitely large, the energy $E$ of the system is given by

\begin{equation}
E={1\over 2}g_1|\Psi_{\alpha}^{\ast}\Psi_{\alpha}|^2+{1\over 2}g_2|\Psi_{\alpha}\Psi_{\alpha}|^2+i\varepsilon_{\alpha\beta\gamma}\omega_{L\gamma}\Psi_{\alpha}^{\ast}\Psi_{\beta}.
\end{equation}
The minimization of $E$ is achieved by 

\begin{equation}
{\vec \Psi}^{(0)}=\sqrt{{n_0\over 2}}(1,i,0) \ \ \ {\rm for}  \ \  \ n_0g_2>-\omega_L
\end{equation}
with $\mu=n_0g_1-\omega_L$.
The order parameter is intrinsically complex and breaks time reversal symmetry. $n_0={\vec \Psi}^{\ast}\cdot{\vec \Psi}$ is the density of the condensate. The magnetic moment ${\bf M}\propto i{\vec \Psi}\times{\vec \Psi}^{\ast}$ whose direction is parallel to the $z$-axis is spontaneously induced in this ground state. This is understandable because $g_2>0$ or $g_s<0 $ implies the ferromagnetic interaction for the spin exchange channel.

For the opposite case, the minimization is achieved by 

\begin{equation}
{\vec \Psi}^{(0)}={1\over\sqrt2}(a,ib,0) \ \ \ {\rm for}  \ \  \ \ n_0g_2<-\omega_L
\end{equation}
where
\begin{equation}
{a \atopwithdelims() b}={\sqrt{n_0\over 2}}\left(\sqrt{ 1+{\omega_L\over n_0|g_2|}}\pm\sqrt{1-{\omega_L\over n_0|g_2|}}\right)
\end{equation}
with $\mu=n_0(g_1+g_2)$. When $B=0$ or $(\omega_L=0)$,
the ground state ${\vec \Psi}^{(0)}={\sqrt n_0}(1,0,0)$ is characterized by a real order parameter, thus it has no spontaneous moment because time reversal symmetry is not broken.

Since the original continuous gauge and spin space rotation symmetries are now broken, the Goldstone modes must exist. Note that the complex order parameter manifold is six-dimensional corresponding to the number of the degrees of freedom in the present system.
In order to derive the collective modes, we consider the equations of motion for small deviation from the ground state: $\Psi_{\alpha}=\Psi_{\alpha}^{(0)}+\delta\Psi_{\alpha}$, which are valid up to linear in $\delta\Psi_{\alpha}$. 
For $n_0g_2>-\omega_L$, the equations of motion for $\delta\Psi_x$ and $\delta\Psi_y$ are coupled, that is,

\begin{eqnarray}
i{\partial\over \partial t} \delta\Psi_x=\{\omega_L+\varepsilon_k+n_0({g_1\over 2}+g_2)\}\delta\Psi_x+{1\over 2}n_0g_1\delta\Psi_x^{\ast}\nonumber \\
+i\{\omega_L+n_0(-{g_1\over 2}+g_2)\}\delta\Psi_y +{i\over 2}n_0g_1\delta\Psi_y^{\ast} \\
i{\partial\over \partial t} \delta\Psi_y=\{\omega_L+\varepsilon_k+n_0({g_1\over 2}+g_2)\}\delta\Psi_y-{1\over 2}n_0g_1\delta\Psi_y^{\ast}\nonumber  \\
-i\{\omega_L+n_0(-{g_1\over 2}+g_2)\}\delta\Psi_x +{i\over 2}n_0g_1\delta\Psi_x^{\ast} 
\end{eqnarray}
where $\varepsilon_k={k^2\over 2m}$ and $\delta\Psi_{\alpha}=\delta\Psi_{\alpha}(k,t)$.
These give rise to a $4\times 4$ eigenvalue matrix while that of $\delta\Psi_z$ is detached from those. After a little manipulation, we find from Eqs.(8) and (9) that the density wave mode and the longitudinal spin wave mode are identical, whose dispersion relation is given by

\begin{equation}
\omega=\sqrt{\varepsilon_k(\varepsilon_k+2n_0g_1)}
\cong ck
\end{equation}
with the velocity

\begin{equation}
c^2={n_0g_1\over m}
\end{equation}
and another mode,

\begin{equation}
\omega=\varepsilon_k+2(\omega_L+n_0g_2).
\end{equation}
The equation of motion for $\delta\Psi_z$:

\begin{equation}
i{\partial\over \partial t} \delta\Psi_z=(\omega_L+\varepsilon_k)\delta\Psi_z 
\end{equation}
leads to the transverse spin wave  mode: 

\begin{equation}
\omega=\omega_L+\varepsilon_k. 
\end{equation}
This massive mode (14) is quadratic in $k$, corresponding to the ferromagnetic-like spin excitation.

For $n_0g_2<-\omega_L$ similar coupled equations of motion
give the following collective modes:

\begin{equation}
\omega^2=\varepsilon_k^2+n_0g_1\varepsilon_k\pm\varepsilon_k\sqrt{n_0^2g_1^2+4g_2(g_1+g_2)\left (n_0^2-{\omega_L^2\over g_2^2}\right )} 
\end{equation}

\begin{equation}
\omega^2=\omega_L^2+\varepsilon_k^2+2n_0|g_2|\varepsilon_k.
\end{equation}
The former is  collective mode coupled with the number density and the longitudinal spin density in the presence of the external field. The latter is the transverse spin wave. When $B=0$, these are decoupled to yield the density wave mode:

\begin{equation}
\omega=\sqrt{\varepsilon_k^2+2\varepsilon_kn_0(g_1+g_2)}\cong ck
\end{equation}
with the zero-sound velocity

\begin{equation}
c^2={n_0(g_1+g_2)\over m}
\end{equation}
and two spin wave modes:

\begin{equation}
\omega= \sqrt{\varepsilon_k^2+2\varepsilon_kn_0|g_2|}\cong ck
\end{equation}
with the spin wave velocity

\begin{equation}
c^2={n_0|g_2|\over m}.
\end{equation}
The former (17) corresponds  to the usual Bogoliubov zero-sound phonon mode while the latter (19) is the spin wave. Both are linear in $k$ in the long wave length limit. It is seen from (18) that for the system to be stable $g_1+g_2>0$ is necessary. The order of magnitudes of the zero-sound velocity and spin wave one are same. Since the former is measured by Andrews {\it et al.}\cite{andrews} on $^{23}$Na atom gas of a single component BEC, which is an order of a few cm/sec and analyzed by the Bogoliubov framework\cite{zaremba,isoshima,pethick}. The known scattering length ($a$=2.75nm) turns out to accurately explain the condensate density dependence of the zero sound velocity, leading mutual reliability between theory and experiment. Here by the same token the measurement of the spin wave velocity may yield a direct estimate for the other missing important parameter $g_2$, which determines the physics of the system
in a fundamental way.

So far we only consider the spatially uniform case. We can expect rich textural spatial structures  or topological defect structures in this multi-component BEC system which is analogous to superfluid $^3$He\cite{volhart}.
In order to help establishing this analogy which may facilitate our understanding, it is useful to adopt the following representation: Let us introduce ${\vec \Psi}={1\over {\sqrt2}}({\bf m}+i{\bf n})$ with ${\bf m}$ and ${\bf n}$ real unit vectors and define ${\bf l}=\bf m\times\bf n$. The three real vectors $({\bf m},{\bf n},{\bf l})$ form a triad and ${\bf l}$ points to the magnetic moment ${\bf M}$. Starting with Eq.(1) we can derive alternative equations of motion in terms of this triad:

\begin{eqnarray}
& & {\partial\over \partial t}n_0=-\nabla_ij_{\rho }^i\\
{\partial\over \partial t} ({\bf m}^2-{\bf n}^2)&=&4\{n_0(g_1+g_2)-\mu\}({\bf m}\cdot{\bf n})\\
& &-2({\bf m}\cdot\nabla^2{\bf n}+{\bf n}\cdot\nabla^2{\bf m})\nonumber\\
{\partial\over \partial t} ({\bf m}\cdot{\bf n})&=&-\{n_0(g_1+g_2)-\mu\}({\bf m}^2-{\bf n}^2)\\
& &-{\bf n}\cdot\nabla^2{\bf n}+{\bf m}\cdot\nabla^2{\bf m}\nonumber\\
{\partial{\bf m}\over \partial t}\cdot{\bf n}-{\bf m}\cdot{\partial{\bf n}\over \partial t}&=&-2\omega_L{\bf l}\cdot{\bf z}+2(n_0g_1-\mu)n_0\\
& &+{1\over 2}g_2({\bf m}^2-{\bf n}^2)^2-
{\bf m}\nabla^2{\bf m}-
{\bf n}\nabla^2{\bf n}\nonumber\\
{\partial\over \partial t} \bf l&=&\omega_L{\bf l}\times{\bf z}-\nabla_i{\bf j}_s^i
\end{eqnarray}
where $n_0={1\over 2}({\bf m}^2+{\bf n}^2)$, the number  current

\begin{equation}
j_{\rho}^i={\bf m}\cdot\nabla_i{\bf n}-{\bf n}
\cdot\nabla_i{\bf m}
\end{equation}

\noindent
and the spin current

\begin{equation}
{\bf j}_s^i=(\nabla_i{\bf m})\times{\bf m}+(\nabla_i{\bf n})\times{\bf n}.
\end{equation}

\noindent
It should be noted that Eq. (25) does not explicitly include the interaction constants $g_1$ and $g_2$. When the system is uniform, the $l$ vector precesses with the Larmor frequency about the equilibrium position.

\begin{figure}
\ 
\hspace{1.5cm}
\epsfxsize=5.5cm
\epsfbox{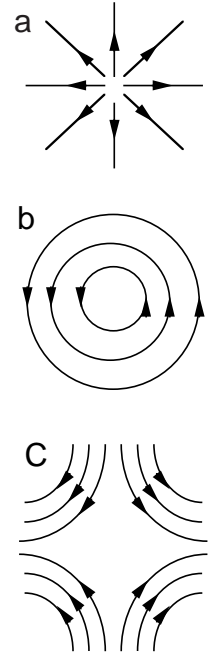}
\caption{
 Possible two dimensional types of the $l$ vector texture.
 (a) radial disgyration, (b) circular disgyration and  (c) cross disgyration 
}
\label{fig:1}
\end{figure}

The $l$-vector field could form various types of the spin texture\cite{maki}, depending on the situation as in superfluid $^3$He such as the radial, circular and cross disgyrations as shown in Fig.1. The last one is particularly interesting because in this configuration the direction of the $l$-vector field coincides with that of the quadrupole magnetic field produced by an optically plugged quadrupole coil. When releasing a BEC  system confined magnetically by the quadrupole trap, this cross disgyration texture may appear as a natural equilibrium configuration.
Since the order parameter of this texture is described by $f(r)(\sin\phi{\bf x}+\cos\phi{\bf y}+i{\bf z})$ in terms of cylindrical coordinate $(r,\phi,z)$, the $x$- and $y$-components exhibit the circulation $\pm1$, implying that this is a kind of the vortex structure. 

It should be noted that the earth field ($\sim$0.1G) which amounts to the Zeeman energy $\sim10\mu K$ is larger than the BEC transition temperature $T_B\sim 1\mu K$, meaning that it cannot be neglected. The order of magnitude of the permissible residual field after shielding the earth field is estimated by equaling the characteristic condensation energy $gn_0\sim{a\over r_0}T_B$ ($r_0$ average interparticle distance $\sim 1000\AA$ and the typical scattering length $a\sim 100\AA$) to the Zeeman energy $\gamma_{\mu} B_c$, resulting in $B_c\sim10^{-3}$G. This size of magnetic shielding is quite feasible.

The best way to detect the spin excitations is of course nuclear 
magnetic resonance experiment, but in view of the small number of the condensed atoms trapped ($\sim10^6$) it may not be possible to 
perform such an experiment. An optical method utilizing the 
Faraday rotation effect of the polarization change due to the magnetization in a medium may be an alternative. We also notice that except for the case of $g_2<0$ under no applied field, all the spin wave excitations are coupled with the density wave excitations, which can be observed as a phase contrast image.

In conclusion, we have derived the coupled non-linear Gross-Pitaevskii equation (3) for BEC with spin degrees of freedom, demonstrating that the vectorial BEC is much richer internal structures than the spin frozen one-component scalar BEC system. In the ground states both for $g_2>0$ and $g_2<0$ not only gauge symmetry but also spin rotation symmetry are broken. The low-lying collective modes, including the gapless Goldstone modes are analytically obtained. In order to help facilitating discussions on spatio-temporal behavior of the multi-component BEC basic equations of motion (21)$\sim$(27) for a triad $(\bf l, \bf m, \bf n)$ are constructed. The detailed study of it belongs to a future work. We propose an experimental setup to create a novel vortex by releasing BEC from a quadrupole trap.

We thank D. Stamper-Kurn for useful communication
on their ongoing experiment and T. Yabuzaki and Y. Takahashi for useful discussions.

\end{document}